\documentstyle[prl,twocolumn,graphicx,aps,amssymb]{revtex}
\begin{document}
\draft
\twocolumn[\hsize\textwidth\columnwidth\hsize\csname    
@twocolumnfalse\endcsname                               
\begin{title} {\Large \bf
Long Range Order at Low Temperature in  Dipolar Spin Ice}
\end{title} 
\author{
Roger G. Melko$^1$,
Byron C. den Hertog$^1$, and
Michel J. P. Gingras$^{1,2}$
}
\address{$^1$Department of Physics, University of Waterloo, Waterloo,
Ontario, N2L 3G1, Canada}
\address{$^2$Canadian Institute for Advanced Research,
                     180 Dundas Street West,  Toronto, Ontario,  M5G 1Z8, Canada}
\date{\today} 
\maketitle 
\begin{abstract}
Recently it has been suggested that long range magnetic dipolar interactions are
responsible for spin ice behavior in the Ising pyrochlore
magnets ${\rm Dy_{2}Ti_{2}O_{7}}$ and ${\rm Ho_{2}Ti_{2}O_{7}}$.
We report here
  numerical results on the low temperature properties of the dipolar
spin ice model, obtained via a new loop algorithm which greatly improves
the dynamics 
at low temperature. We recover the previously reported missing
entropy in this model, and find a first order transition to a long range ordered phase with
zero total magnetization at very low temperature. We  discuss the relevance of
these results to
${\rm Dy_{2}Ti_{2}O_{7}}$ and ${\rm Ho_{2}Ti_{2}O_{7}}$.

\end{abstract}

\vskip2pc] 

\begin{figure}
\begin{center}
\includegraphics[width=7.cm]{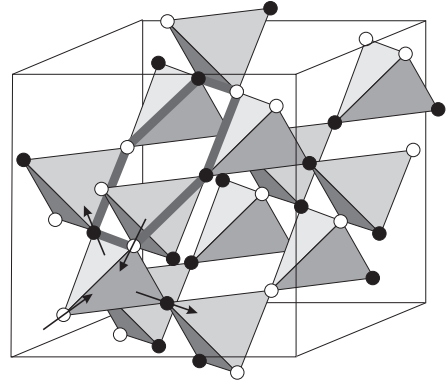}
\vspace{-5mm}
 \caption{The lower left `downward'
tetrahedron of the pyrochlore lattice
 shows Ising spins (arrows) whose axes meet at the middle of the
tetrahedron. For clarity,  black and white circles on the lattice points
denote other spins. White represents a spin pointing into a downward
tetrahedron while black is the opposite. The entire  lattice is shown in
an ice-rules state (two black and two white sites for every tetrahedron).
The hexagon (thick gray line) shows a minimal loop move, which corresponds to
reversing all colors (spins) on the loop to produce a new ice-rules state.}
 \label{fig1}
\end{center}
 \end{figure}

In frustrated magnetism, the term
{\it spin ice}
was recently coined by Harris and 
coworkers \cite{Harris} to 
describe the  analogy that exists between the statistical physics 
of  certain geometrically frustrated Ising pyrochlore magnets, and proton 
ordering  in the hexagonal phase of ice (${\rm I_{h}}$) \cite{Bernal,Pauling,Anderson}. 
For the Ising pyrochlore systems 
${\rm Ho_{2}Ti_{2}O_{7}}$ and ${\rm Dy_{2}Ti_{2}O_{7}}$, the ${\rm Ho^{3+}}$ 
and ${\rm Dy^{3+}}$ rare earth magnetic
moments reside on a   network of corner sharing tetrahedra 
(Fig. \ref{fig1}). Each moment is 
forced by single-ion 
anisotropy to lie along the axis joining the centers of the two tetrahedra 
that it belongs to\cite{Harris,Ramirez}. For  a simple theoretical model 
considering only nearest neighbor ferromagnetic (FM) exchange, the 
groundstate is 
macroscopically 
degenerate, but is required to  have two moments pointing in and two 
pointing out of 
every tetrahedron,  a constraint that  maps exactly the two short and 
long proton bonds and  the ice-rules for their arrangement in  
${\rm I_{h}}$ \cite{Bramwell,Harris2}.
This nearest neighbor FM  model shows no ordering
and is characterized by a broad Schottky-like  peak in the magnetic
specific heat\cite{Harris2}.

Both ${\rm Ho_{2}Ti_{2}O_{7}}$ \cite{Harris,Bramwell2}
 and ${\rm Dy_{2}Ti_{2}O_{7}}$  
	\cite{Ramirez,denHertog}
show qualitative properties roughly consistent with the basic spin ice picture
of the simple nearest neighbor FM model
\cite{Bramwell,Harris2}.
However, it has been 
 shown recently that rather than nearest neighbor FM exchange, it is 
surprisingly the large dipolar interaction present in these materials that is 
 responsible for their spin ice
 behavior\cite{Bramwell2,denHertog,Gingras,note2,note}. 
 For a model which we call \mbox{{\it dipolar spin ice}}, 
 with the long range nature of the dipolar interaction properly handled using
 Ewald summation techniques,
 numerical results 
 show a lack of magnetic  ordering down to 
very low temperatures\cite{denHertog}.
Furthermore, 
the dipolar spin ice model agrees
quantitatively very well
with specific heat data for ${\rm Dy_{2}Ti_{2}O_{7}}$ \cite{Ramirez} 
and  ${\rm Ho_{2}Ti_{2}O_{7}}$ \cite{Bramwell2}, as well as neutron scattering 
measurements on the latter material\cite{Bramwell2}.
In other words, while the simple nearest neighbor FM model provides a 
simple and qualitative understanding of the spin ice phenomenon,
there is now strong evidence that the dipolar spin ice model with
its long range dipolar interactions provides a quantitatively
accurate description of experimental results on real materials 
\cite{Bramwell2,denHertog}.
\vspace{-4mm}
\vspace{-4mm}
\noindent 
As in the case of ${\rm I_{h}}$, for dipolar spin ice  it is 
unclear whether  the long range interaction should cause 
the absence of ordering (and nonzero entropy) down to zero temperature. 
The dipolar interaction is itself FM at nearest neighbor, and  is thus prone 
to spin ice correlations. However, {\it a priori} one might expect that
 its longer range component should lift the nearest neighbor degeneracy and 
induce the selection of an ordered state within the ice-rules manifold.
We show in this work that this is precisely the case.
Specifically, 
the dipolar spin ice model with long range interactions
does possess a unique groundstate
(apart from trivial global symmetry operations)
which develops at very 
low temperature. However,
for local dynamical processes (such as single spin 
fluctuations), the development of this ground state 
is completely dynamically 
inhibited. As we discuss 
below, this occurs because of high energy barriers separating
quasi-degenerate  ice-rules states. This allows paramagnetic spin 
ice behavior to occur despite no special spin or space
 symmetry  in the 
system which would a priori prevent magnetic ordering.
In this paper
we explore the low temperature ordering  properties  of dipolar
spin ice by taking    advantage of  `loop moves' incorporated into a standard
Metropolis Monte Carlo algorithm,  a method  considered  previously in
 the context of two-dimensional square ice models\cite{Barkema}.
Such moves allow us to explore  degeneracy lifting effects within the ice-rules
manifold  in an efficient manner, something which is not possible via single
spin flip fluctuations. We present here strong  numerical evidence
 for  a first order phase transition at extremely low temperature in the
dipolar spin ice model in zero field that recovers the entire low temperature residual 
magnetic entropy of the system.

For the pyrochlore lattice with Ising spins defined by local axes, 
the Hamiltonian with nearest 
neighbor exchange and long range dipolar interactions is
\cite{Bramwell2,denHertog,Gingras}:
\begin{eqnarray}
\label{eqn1}
H&=&-J\sum_{\langle ij\rangle}{\bf S}_{i}^{z_{i}}\cdot{\bf S}_{j}^{z_{j}}
\nonumber \\
&+& Dr_{{\rm nn}}^{3}\sum_{i>j}\frac{{\bf S}_{i}^{z_{i}}\cdot{\bf
S}_{j}^{z_{j}}}{|{\bf r}_{ij}|^{3}} - \frac{3({\bf S}_{i}^{z_{i}}\cdot{\bf r}_{i
j})
({\bf S}_{j}^{z_{j}}\cdot{\bf r}_{ij})}{|{\bf r}_{ij}|^{5}} \; ,
\end{eqnarray}
where the spin vector ${\bf S}_{i}^{z_{i}}$ labels the Ising moment of
magnitude $|S|=1$ at lattice site $i$ and {\it local}  Ising $[111]$ 
 axis $z_{i}$ discussed earlier. Here
$J$ represents the
 exchange energy and $D=(\mu_{0}/4\pi)g^{2}\mu^{2}/r_{nn}^{3}$. However,
because of the local Ising axes, 
 the effective nearest neighbor 
energy scales are $J_{\rm nn}\equiv J/3$ and $D_{\rm nn}\equiv 5D/3$. 

As described in Ref. \cite{denHertog}, the long range nature of the dipolar 
interactions can be handled conveniently by the Ewald method. In that 
work, extensive numerical analysis via single spin flip Monte Carlo simulations
 found no evidence of a transition to long range order.  Rather, 
short range order dominated by ice-rules correlations was observed down to low
 temperatures, similar to that found in the 
nearest neighbor FM model\cite{note}. 

Qualitatively, the dynamics of both models appear to be very similar. 
As the temperature is lowered, significant thermal 
barriers are created by the energy cost involved in fluctuating {\it out} of 
the 
ice-rules manifold. With single spin flips, fluctuations {\it between} states 
{\it within} the ice-rules manifold are also reduced, as it is 
impossible to do so without first breaking the two-in/two-out ice-rules.
 Such thermal barriers produce non-trivial and extremely slow dynamics. 
If a unique groundstate exists within the plethora of ice-rules states 
($\sim (3/2)^{N/2}$) of the dipolar spin ice model (Eq. 1), 
these thermal barriers make the  probability of 
reaching it in a numerical simulation using conventional 
spin flips  exceedingly  small. Consequently, the question concerning
the nature of the groundstate becomes difficult to  answer using standard 
numerical techniques, and a different procedure must be applied\cite{Barkema}.
Since we found in Ref.\cite{denHertog} that long range dipolar interactions
give rise to spin ice behavior, we take as a starting point for identifying
the low energy excitations (quasi zero modes) of Eq. (1)
the exactly degenerate ice-rules state.
This is entirely analogous to the approach taken in considering the so-called
`energetic ice models' in two-dimensional square ice models~\cite{Barkema}.
\begin{figure}
\begin{center}
\rotatebox{90}{\includegraphics[width=6cm]{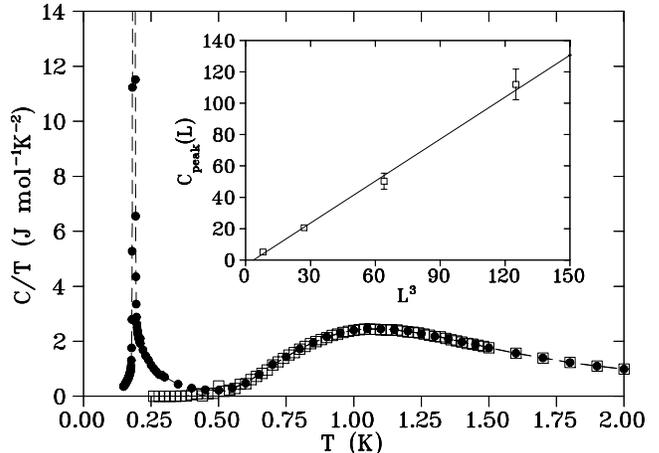}}
 \caption{Specific heat data from simulations using single spin flips (squares) 
and combined single spin flips and loop moves (filled 
circles). Interaction parameters are   ${\rm Dy_{2}Ti_{2}O_{7}}$ values
 $J_{\rm nn}=-1.24 \; {\rm K}$ and $D_{\rm nn}=2.35 \;{\rm K}$ 
(as in Ref. [9]). 
Inset: Finite size scaling of the specfic heat peak height as a function of
 system size $L=2,3,4,5$. The scaling behavior 
$C_{\rm peak}(L) =a + bL^3$ is consistent with that
expected for a first order phase transition.}
 \label{fig2}
\end{center}
 \end{figure}
\vspace{-4mm}
\noindent In Fig. \ref{fig1} we denote each site of the
pyrochlore lattice by a white or black circle which represents a spin 
pointing into or out of a `downward' facing tetrahedron, respectively. 
In this particular example, the spin configuration shown  forms an ice-rules 
state that can be transformed into
 another ice-rules state by reversing all the colors (spins) on the loop denoted by the 
gray hexagon. In general, six spins form the shortest loop, 
while larger  loops are also possible.  A loop can be constructed by
 simply  choosing a starting lattice site 
and tracing out a closed path that 
involves tetrahedra which have exactly two spins on the path (see Fig. 1). 
Furthermore,  each pair 
of spins  which are neighbors on the path
are such that one is pointing 
 into and the other pointing out of their shared tetrahedron. As seen 
 in Fig. \ref{fig1}, such a loop is constructed of 
alternating black and white circles. 

For our numerical study of the dipolar spin ice 
model, this type of `loop move' was utilized in conjunction with 
conventional single spin flip dynamics. Specifically, such loops are
identified by  allowing a wandering path to form a loop whenever  
it encounters any previously visited 
site and ignoring any `dangling' spins in the path.  This allows for 
a large number of
short loops to be created, with an average length 
 that tends to a finite
 value as  the system size is increased. 
As explained above for the dipolar system, such `loop reversal'
moves are not true zero modes, 
but involve a small gain or lowering of the energy (small compared to
$J_{nn}+D_{nn}$) which is handled by 
a standard Metropolis algorithm\cite{note4}.

Our numerical simulations for the dipolar spin ice model were carried out on
 system sizes up to 
2000 spins (of cubic unit cell length L=5) with  $O(10^{5})$ 
spin flips per spin and $O(10^{5})$ loop moves. 
For all interaction parameters $J_{\rm nn}$  and $D_{\rm nn}$ 
which show spin ice behavior  using single spin flip dynamics only  
($J_{\rm nn}/D_{\rm nn} \gtrsim -0.91$)\cite{denHertog},
 we find that the acceptance ratio of the loop moves 
increases at low temperature as the system enters the spin ice regime, 
before dropping to zero just below the temperature at which  the system  
appears to undergo a very sharp first order phase 
transition to a long range ordered state obeying the ice-rules.

In Fig. \ref{fig2}  we present specific heat data obtained for a system
with interaction parameters $J_{\rm nn}$  and $D_{\rm nn}$ identified in Ref.
\cite{denHertog} for the spin ice material ${\rm Dy_{2}Ti_{2}O_{7}}$. 
Using a single spin flip Monte Carlo algorithm, 
spin ice correlations develop 
over a large 
temperature regime (signified by the broad peak
around 1.1 K), before the system dynamically slows down into a disordered 
ice-rules state at low temperature. Using the loop algorithm in combination
 with single spin flips, the higher temperature
data is reproduced before a very sharp transition is observed at 
$T_{c}\simeq 0.18\; {\rm K}$, with  a latent heat observed
at the transition. The energy probability distribution displays a double-peak
feature in a narrow temperature region close
to $T_c$, another indicator that the transition is first order.
To assess in a more quantitative way the
nature of the phase transition, a finite-size scaling study was done (see inset
of Fig. 2).
Because of the extremely sharp nature of the specific heat at $T_{c}$,
the method of slowly cooling in a Monte Carlo simulation with discrete
temperature steps could not give sufficiently accurate data to resolve
$C_{peak}$
within reasonable computer time.
To avoid this problem, simulations were performed in a multicanonical
ensemble \cite{Hansmann}
at a single temperature near $T_{c}$.  This data was
then re-weighted using Ferrenberg and Swendsen's technique~\cite{Ferrenberg},
which
allowed us to obtain the appropriate thermodynamic quantities 
to any
degree of temperature resolution required.

The ordered phase is similar to that found in the
order by disorder transition in the 
antiferromagnetic FCC Ising model\cite{Wengel}. In that frustrated system,  an ordering of 
antiferromagnetically stacked FM planes is found. 
For  the dipolar spin ice system considered here, the ordering vector
 ${\bf q}$ lies parallel to one of the cubic axes directions, specifically  
${\bf q}=(0,0,2\pi/a)$ or its starred directions. 
To construct the ordered state, first consider a starting tetrahedron with
its six possible ice-rules states. For a given ordering vector ${\bf q}$, 
this tetrahedron selects one of the four possible spin configurations
(two independent configurations and their spin-reversals, ${\bf S}_i 
\rightarrow -{\bf S}_i$), with a total magnetic moment for the tetrahedron
perpendicular to ${\bf q}$.
The entire ordered state may then be described  by  planes 
(perpendicular to ${\bf q}$) of such tetrahedra.
 The wavelength defined by  ${\bf q}$  physically  corresponds to antiferromagnetically 
stacked planes of tetrahedra, where  a given plane has  
tetrahedra of opposite configuration to the plane above and below it. 
In Fig. 3 we show one such groundstate with ordering vector ${\bf q}=(0,0,2\pi/a)$.

The transition to such  a groundstate structure can be characterized  by the 
multi-component  
order parameter
\begin{equation}
{ \Psi}_{\alpha}^{m} =\frac{1}{N}\left|\sum_{j=1}^{N/4}\sum_{a=1}^{4}
\sigma^{j}_{a} 
{\rm e}^{i\phi^{m}_{a}}{\rm e}^{i{\bf q}_{\alpha}.{\bf r}_{j}}\right| \; .
\end{equation}
 Such a labeling is natural given that the pyrochlore lattice can be viewed 
 as an FCC 
lattice with a `downward' tetrahedral basis (see Fig. 1).
Thus $j$ labels the FCC lattice points of the pyrochlore 
lattice, and the index $a$ sums over the four spins 
comprising the basis attached to each $j$. The index
$\alpha$ labels the three possible symmetry related  ${\bf q}$ ordering 
vectors. For a given ${\bf q}_{\alpha}$, 
as described above, there are two  ice-rules configurations and their reversals
which can each form a groundstate. Thus  $m=1,2$ labels these possibilities 
with the phase factors $\{\phi^{m}_{a}\}$ describing the  given configuration 
$m$. Each Ising variable $\sigma_a^j$ has value 1 (-1) when a spin points into 
(out of) its downward tetrahedron labeled by $j$.

As written in Eq.(2), ${\Psi}_{\alpha}^{m}$ has six degenerate
components, each of which can take on a value between 0 and 1.
Upon cooling through the transition, the system selects a unique
ordered configuration, causing the corresponding component of
${\Psi}_{\alpha}^{m}$
to rise to unity and all others to fall to zero.
The component selected by the ordering is equally likely to be any one
of the six.
Fig. 3 is a plot of $\left<\Psi \right>$ for two system sizes, where
$\left<\Psi \right>=\sqrt{\sum_{m=1}^{2}\sum_{\alpha =1}^{3}\left(
{\Psi}_{\alpha}^{m} \right)^{2}}
$
is the magnitude of the multi-component order parameter.
For $T<T_{c}$ the two lattice sizes produce
identical order parameters. By contrast, $\left<\Psi \right>$ for 
the smaller lattice shows a somewhat more 
pronounced rounding near $T_{c}$, and an
increased
residual value for large $T$.  These results show a clear discontinuity
of the
order parameter at $T_c$, and hence a first order
transition to the long range ordered phase we have identified.

For all values of $J_{\rm nn}/D_{\rm nn}$ within the 
dipolar spin ice regime \cite{denHertog}, we find a low 
temperature phase transition to the state discussed above. The 
transition is driven by the long range dipolar interactions and,
therefore, $T_c\sim 0.18$K  is independent of the strength of the
nearest neighbor exchange $J$ ($T_c/D_{nn} \sim 0.08$). 
The observation of a finite 
ordering temperature using the  algorithm presented 
here demonstrates  that long range dipolar interactions between
Ising spins on the pyrochlore lattice have no special exact symmetry that 
allow   a  macroscopically degenerate ground state.
 This conclusion is also suggested  within a mean field 
analysis\cite{denHertog,Gingras}, which shows that as the truncation of
 long range 
dipolar interactions is pushed out to further distances (up to $10^{4}$ 
nearest neighbors), the maximal eigenvalues of the
normal mode spectrum become only {\it quasi-degenerate} throughout the 
Brillouin zone, as opposed to the completely flat spectrum (and  macroscopic 
degeneracy) we find for the
nearest neighbor spin ice model \cite{Gingras}. Furthermore, 
 the  quasi-degenerate eigenvalues of the mean field theory have 
a very weak dispersion which is maximal at the FCC zone boundary and,
therefore, predicts the same 
ordering wavevector ${\bf q}$ found here.

\vspace{-5mm}
 \begin{figure}
\begin{center}
\includegraphics[width=7.cm]{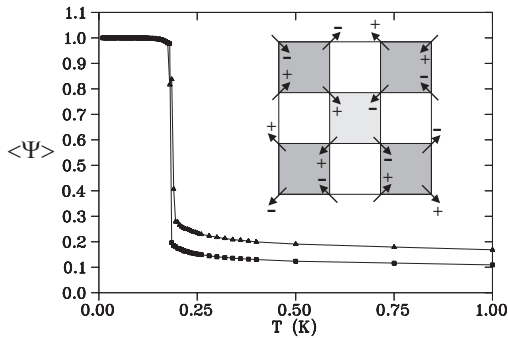}
\vspace{-1mm}
 \caption{ 
 Temperature dependence of the order parameter $\left<\Psi
\right>$
defined above for system sizes L=3 (triangles) and L=4
(squares).  Inset: The ${\bf q} = (0,0,2\pi /a)$ groundstate
projected
down the z axis.  The four tetrahedra 
making up the cubic unit cell 1
appear as dark grey squares.  The light grey square does not represent
a tetrahedron, however its diagonally opposing spins occur in the same
plane.
The component of each spin parallel to the z axis is
indicated by a $+$ and $-$ sign.}
\label{fig3}
\end{center}
 \end{figure}

\vspace{-5mm}

The  question   remains as to what extent  our 
conclusions apply to the real spin ice materials ${\rm Ho_{2}Ti_{2}O_{7}}$\cite{Harris} and 
${\rm Dy_{2}Ti_{2}O_{7}}$\cite{Ramirez}. The dipolar spin ice model may be  an accurate 
description of these materials even  at extremely low temperatures, while
it is also   possible that  
perturbations, $H^{\prime}$, exist  beyond Eq.~\ref{eqn1}  which 
could  induce another type of groundstate selection.
Similar to  ${\rm I_{h}}$ however,
irrespective of the origin of any ordering,   its  actual  observation  may 
depend critically on
the dynamical behavior of the materials. The inability of single spin 
fluctuations to  connect different ice-rules states in phase space 
shows that at low temperatures 
relaxation via local dynamics is extremely slow. 
For both  ${\rm Ho_{2}Ti_{2}O_{7}}$ and  ${\rm Dy_{2}Ti_{2}O_{7}}$, the 
transition temperature for the ordered phase observed in our simulations is 
well below the temperature at which single
 spin fluctuations over extended length scales (and out of the ice-rules 
manifold)
 are thermally frozen out. Thus, while theoretically  an ordered phase induced
 by long range dipolar interactions 
between Ising spins on the pyrochlore lattice does exist, its experimental
observation will depend acutely on the dynamical processes of the  real  
materials. Furthermore, one requires that perturbations $H^{\prime}$ are negligible, ie. that 
$H^{\prime}/D_{\rm nn}\lesssim T_{c}/D_{\rm nn}\lesssim 0.08$.

In conclusion, we predict that in the
dipolar spin ice model, which is in quantitative agreement with experimental
 data on real systems in the temperature regime
investigated so far \cite{Bramwell2,denHertog},
  a very low temperature transition  
to a zero total moment structure exists with recovery of all residual entropy. 
However, it is unlikely that such a phase can be arrived at via conventional 
local dynamics. These results suggest  that Ising pyrochlore magnets
 with long range dipolar 
interactions provide an even deeper analogy with the proton ordering in 
hexagonal ice water ${\rm I_{h}}$ than previously suggested.

We  thank   S. Bramwell and P. Holdsworth  for  useful discussions. 
R.M.  acknowledges financial support from NSERC of Canada.
 M.G. acknowledges financial support from NSERC,
Research Corporation  and the Province of Ontario.

%
%

%
%


\begin{references}
\bibitem{note2}
The nearest neighbor exchange in ${\rm Ho_{2}Ti_{2}O_{7}}$ and 
${\rm Dy_{2}Ti_{2}O_{7}}$  is 
antiferromagnetic, which by 
itself would trivially cause long range N\'{e}el order. 
See Refs. \cite{Bramwell2,denHertog}
\bibitem{note}
There are subtle differences between the correlations in the dipolar spin ice 
model and
those found in the  nearest neighbor spin ice model.
For example,
the dipolar spin ice model reproduces a larger amount of detail
in the structure factor of
${\rm Ho_{2}Ti_{2}O_{7}}$. See Ref. \cite{Bramwell2}.
\end{references}
\end{document}